# Impact of Chemical Clustering on the Structural, Topological and Functional Properties of $Ba(Zr_xTi_{1-x})O_3$: An Atomistic Simulation Study

M. Baldassin, R. Machado, M. Sepliarsky and M.G. Stachiotti

*Instituto de Física Rosario, Universidad Nacional de Rosario, 27 de Febrero 210 Bis, (2000) Rosario, Argentina.*

Barium zirconate titanate, $Ba(Zr_xTi_{1-x})O_3$ (BZT), is a leading lead-free candidate for high-performance electronic components due to its highly tunable ferroelectric-to-relaxor transition. In this study, we employ molecular dynamics simulations to investigate the effects of local chemical clustering on the structural and functional properties of BZT in both ferroelectric (x = 0.2) and relaxor (x = 0.4) regimes. Crucially, we demonstrate that Zr clustering induces a significant enhancement of local polarization across both compositions. Rather than a purely volume-driven effect, this enhancement is fundamentally governed by a spatial redistribution of local structural phases (rhombohedral, orthorhombic, and tetragonal); Zr segregation forms larger, continuous Ti-rich regions that expand highly polar rhombohedral domains and foster cooperative dipolar alignment. This cooperative coupling thermally stabilizes the local polar order, systematically elevating the Curie, maximum permittivity, and Burns temperatures. Focusing on the relaxor composition (x = 0.4), chemical segregation promotes the formation of resilient, swirling polar topological textures—categorized into vortices circulating around Zr-rich clusters and interstitial vortices localized within the Ti-rich matrix. These topological structures act as effective pinning barriers against polarization reversal, driving pronounced electrical hardening with elevated coercive fields. Ultimately, these findings establish a quantitative link between nanoscale compositional heterogeneity, polar domain topology, and macroscopic performance, providing a robust framework for engineering lead-free perovskites via local chemical order control.

*Keywords: Ferroelectrics; Relaxors; Barium titanate; BZT; Atomistic modeling*

## 1. Introduction

Lead-free perovskite oxides have become the focus of intense research driven by environmental regulations and the demand for sustainable electronic components. Within this framework, barium zirconate titanate, $BaZr_xTi_{1-x}O_3$ (BZT), has emerged as a prominent candidate to replace lead-based compositions in dielectric, energy-storage [1] and energy-harvesting applications [2,3]. This solid solution stands out for its remarkable dielectric permittivity, low loss, and the emergence of a highly flexible polar state that can be tailored through chemical substitution.

The BZT phase diagram is remarkably diverse, exhibiting distinct physical regimes governed by the zirconium concentration (x) [4-7]. At low concentrations ($x < 0.11$), the material behaves as a conventional ferroelectric—analogous to $BaTiO_3$—undergoing three well-defined structural phase transitions and establishing long-range ordered domains. In the intermediate range ($0.11 < x < 0.27$), these transitions collapse into a single, diffuse ferroelectric-to-paraelectric transition, characterized by a significant broadening of the dielectric permittivity peak. As the Zr content increases further ($x > 0.27$), BZT transitions into a relaxor state; here, the permittivity maximum ($T_m$) exhibits a hallmark frequency-dependent dispersion driven by local polar order, while the average macroscopic structure remains pseudocubic across the entire temperature range. At the Zr-rich limit ($x > 0.8$), the system eventually enters a dipolar glass state. This intricate phase evolution has been successfully captured by first-principles-based approaches, ranging from effective Hamiltonians [8-13] to atomistic simulations [14]. In the latter work, we introduced a local-phase-based approach to unify the description of the distinct regions of the solid solution—including the ferroelectric, relaxor, and dipolar glass phases—thereby capturing the continuous evolution from one regime to another. These findings demonstrated how atomic-level descriptions drive the complex macroscopic behavior of the BZT system.

Regarding the temperature evolution within the relaxor regime, the high-temperature paraelectric matrix undergoes subtle local changes upon cooling. Traditionally, this process has been interpreted as the nucleation and growth of isolated polar nanoregions (PNRs) that emerge at the Burns temperature ($T_B$), well above the permittivity maximum [15-19]. More recently, however, an alternative viewpoint has emerged, suggesting that this intermediate state may be better described by interconnected, "slush-like" polar configurations characterized by a highly inhomogeneous distribution of continuous local displacements [12,20,21]. Regardless of the framework, the precise nature of these local states at high temperatures and their sensitivity to structural perturbations remain central topics of discussion.

Another critical and highly controversial aspect of BZT is its internal chemical order and the exact mechanism by which local arrangement dictates relaxor behavior. While BZT is frequently modeled as an ideal, statistically random solid solution, advanced local probes have revealed a much more complex microstructural reality. Extended X-ray Absorption Fine Structure (EXAFS) studies carried out by Laulhé et al. [22] demonstrated that to accommodate the severe lattice misfit induced by the larger $Zr^{4+}$ ions within the $BaTiO_3$ matrix—originating from the significant ionic size mismatch ($r_{Zr}/r_{Ti}$=1.18)—the local lattice undergoes significant structural distortions. Crucially, these analyses provided quantitative evidence of chemical heterogeneity, revealing that Zr atoms exhibit a distinct tendency to segregate into local $BaZrO_3$-like nanoregions. For a representative concentration of x = 0.35, and assuming a model of hypothetical spherical inclusions, this segregation corresponds to a mean cluster size containing approximately 27 Zr atoms, yielding a local radius of roughly 7.2 Å [22].

However, the general validity of this segregation hypothesis remains a subject of debate. Subsequent XAFS investigations coupled with Density Functional Theory (DFT) calculations by Levin et al. [23] did not detect definitive evidence of such short-range clustering across intermediate compositions, suggesting instead that BZT can behave as a statistically disordered solid solution where the ionic size mismatch is resolved through uniform internal relaxations. This dispute underscores a broader problem in the physics of relaxor ferroelectrics: understanding the precise correlation between nanoscale chemical order and polar short-range order. In classical lead-based relaxors like $Pb(Sc_{1/2}Nb_{1/2})O_3$ (PSN), the diffuse state is driven by aliovalent cationic disorder that generates long-range random electric fields. In such systems, Molecular Dynamics (MD) simulations by Burton et al. [24] have demonstrated that chemically ordered regions act as islands of enhanced polarization and intense dielectric fluctuations compared to the surrounding disordered matrix, where high local electric fields actively pin cooperative dipole rotations. In BZT, however, electrostatically driven random fields are absent due to the homovalent substitution of $Zr^{4+}$ and $Ti^{4+}$. Instead, the system is governed by random elastic fields originating from the ionic size mismatch, meaning that the link between chemistry and polarity becomes highly localized. While an isolated $Ti^{4+}$ ion within an expanded $BaZrO_3$ host lattice remains centrosymmetric and nonpolar, the statistical emergence or chemical segregation of nearest-neighbor Ti-Ti pairs breaks this local symmetry, forcing polar off-center displacements that serve as the seeds for local polarization [14,23,25]. Consequently, the degree of chemical nanostructuring stands out as the ultimate mechanism modulating the functional and relaxor performance of these lead-free materials. This suggests that, in the final analysis, the actual existence and scale of these segregated nanoregions may fundamentally depend on the specific synthesis and processing conditions of the samples.

In this work, we use Molecular Dynamics (MD) simulations to systematically investigate the effect of Zr segregation on the structural, dielectric, and ferroelectric properties of BZT. We focus on two representative concentrations: x=0.2, corresponding to the diffuse ferroelectric regime, and x=0.4, associated with the conventional relaxor behavior. By modeling different degrees of Zr aggregation—ranging from perfectly random solid solutions to explicitly segregated nano-clusters—we analyze the resulting changes in macroscopic and local polarization, dielectric permittivity, as well as transition and relaxation temperatures. This systematic approach allows us to evaluate how chemical nanostructuring influences the local polar configurations, shedding light on the mechanisms that drive the material's response. Finally, we evaluate the impact of chemical disorder on the stability of topological structures within the polarization field. Our results provide a quantitative link between chemical disorder and the modification of functional properties, offering new insights into the microscopic nature of lead-free relaxor ferroelectrics.

## 2. Computer simulation approach

Atomistic simulations were performed using the Molecular Dynamics (MD) technique as implemented in the DL_POLY package [26]. The interatomic interactions were described by a shell model potential, which effectively accounts for the electronic polarizability of the ions. In this framework, each polarizable ion (Ba, Ti, Zr, and O) is represented by a core containing the nuclear mass and a massless shell, connected by an anharmonic spring. The core-shell separation mimics the dipole moment induced by local electric fields.

The short-range interactions were modeled using a combination of Born-Mayer ($V(r)=Ae^{-r/\rho}$) and Buckingham ($V(r)=Ae^{-r/\rho} - C/r^6$) potentials, while the long-range electrostatic interactions were treated using the Wolf summation method. This approach provides a computationally efficient alternative to the Ewald sum for large-scale systems without sacrificing accuracy in the calculation of Coulombic forces. The potential parameters were optimized against first-principles data, demonstrating their reliability in previous studies where they successfully reproduced the properties of $BaTiO_3$ [27] and $BaZrO_3$ [28] systems, as well as in BZT solid solutions [14] and nanocomposites [29].

The MD simulations were conducted within the constant stress and temperature (N, σ, T) ensemble. In this way, the size and shape of the simulation cell were dynamically adjusted to obtain the desired average pressure. Simulations were carried out in a supercell consisting of 28 x 28 x 28 perovskite unit cells (109,760 atoms), ensuring that the system size was sufficient to capture the formation of polar nanostructures and avoid periodic boundary artifacts. The simulations were conducted at temperature intervals of 10 K, with a time step of 0.4 fs. Each MD run consisted of at least 100,000 time steps (40 ps) for data collection, following a thermalization period of 20,000 steps (8 ps).

To characterize the local polar behavior of the system, we define a local dipole moment vector $\boldsymbol{p}_i = (p_i^x, p_i^y, p_i^z)$ for each perovskite unit cell i centered at the B-site cations (Ti or Zr atoms). The magnitude of the average local polarization $P_L$ is then evaluated as:

$$P_L = \langle [\langle p_i^x \rangle_t^2 + \langle p_i^y \rangle_t^2 + \langle p_i^z \rangle_t^2]^{1/2} \rangle_i \quad (1)$$

where the inner brackets $\langle ... \rangle_t$ denote the temporal average taken over the production MD trajectory, and the outer bracket $\langle ... \rangle_i$ represent the spatial average computed over all equivalent lattice sites. Conversely, the macroscopic polarization vector **P** is calculated as:

$$\mathbf{P} = ( \langle \langle p_i^x \rangle_t \rangle_i , \langle \langle p_i^y \rangle_t \rangle_i , \langle \langle p_i^z \rangle_t \rangle_i ) \quad (2)$$

To model the effect of chemical disorder, we considered three types of zirconium distributions denoted as S0, S50, and S100, representing different degrees of cationic segregation. The S0 configuration corresponds to an ideal random solid solution where zirconium and titanium ions are distributed stochastically across the B-sites of the perovskite lattice. In the S50 model, fifty percent of the zirconium ions are clustered into specific nanoregions while the remainder remains randomly distributed, whereas the S100 model features total segregation with all zirconium ions concentrated into well-defined clusters. The morphology of these aggregates was modeled as spheres with a fixed radius of two lattice constants, resulting in clusters composed of 33 perovskite unit cells. Spatial overlap between adjacent spheres was permitted during the generation of the configurations. This specific dimension was chosen based on estimations from EXAFS studies [22], which indicate that chemical inhomogeneities in BZT typically occur at this nanoscopic scale. This approach allows for the isolation of the impact of chemical arrangement from other structural variables by maintaining a constant cluster volume across the different segregation scenarios.

To gain insight into the microscopic characteristics of the systems and quantify the coexistence of local phases, we employed the unsupervised machine learning method Multi-SOM [30] to classify individual cells into distinct local phase types based on their polarization vectors. Briefly, the method focuses exclusively on polar-active Ti-centered cells, separating polar from nonpolar groups via a data-driven polarization threshold of 13 $\mu C/cm^2$. Subsequently, the three polar phases (tetragonal, orthorhombic, and rhombohedral) are distinguished by using rotationally symmetric normalized dipole components as descriptors, where the components are ordered by magnitude ($p_{max}$, $p_{mid}$, $p_{min}$) and normalized so that $p_{max}$ equals 1. The resulting two-dimensional space of ($p_{mid}$, $p_{min}$) is tessellated into distinct regions: the tetragonal phase is identified by $p_{mid} < 0.5$ and $p_{min} < 0.35$, the orthorhombic phase by $0.5 < p_{mid} \leq 1$ and $p_{min} < 0.35$, and the rhombohedral phase by $0.35 < p_{mid} \leq 1$ and $p_{min} \geq 0.35$. Further methodological details regarding this classification tool can be found in Ref. [14].

## 3. Results and Discussion

### 3.1.Validation of the Model for Random Solid Solution

Before analyzing the effects of chemical segregation, the atomistic model was validated by examining the temperature dependence of the local and macroscopic properties for the random solid solution (S0) in both the diffuse ferroelectric (x = 0.2) and relaxor (x = 0.4) regimes. **Figure 1** displays the temperature evolution of the macroscopic polarization, the average local polarization ($P_L$), and the average lattice parameters for both compositions.

For the x = 0.2 composition, the model successfully reproduces a well-defined ferroelectric-to-paraelectric phase transition, characterized by a sharp drop in macroscopic polarization and a corresponding abrupt change in the lattice parameters. In contrast, at x = 0.4, the net macroscopic polarization remains zero across the entire temperature range, while the unit cell volume exhibits a smooth, continuous evolution. This behavior represents a hallmark of the relaxor state, where the absence of long-range ferroelectric order persists despite the presence of robust local polar order, as evidenced by the thermal evolution of the local polarization insets. These results confirm that our atomistic potential and simulation protocol accurately capture the fundamental physics of BZT across both the conventional ferroelectric and relaxor regimes. A comprehensive exploration of the intricate BZT phase diagram over the full concentration range, obtained via this same atomistic approach, is detailed in Ref. [14].

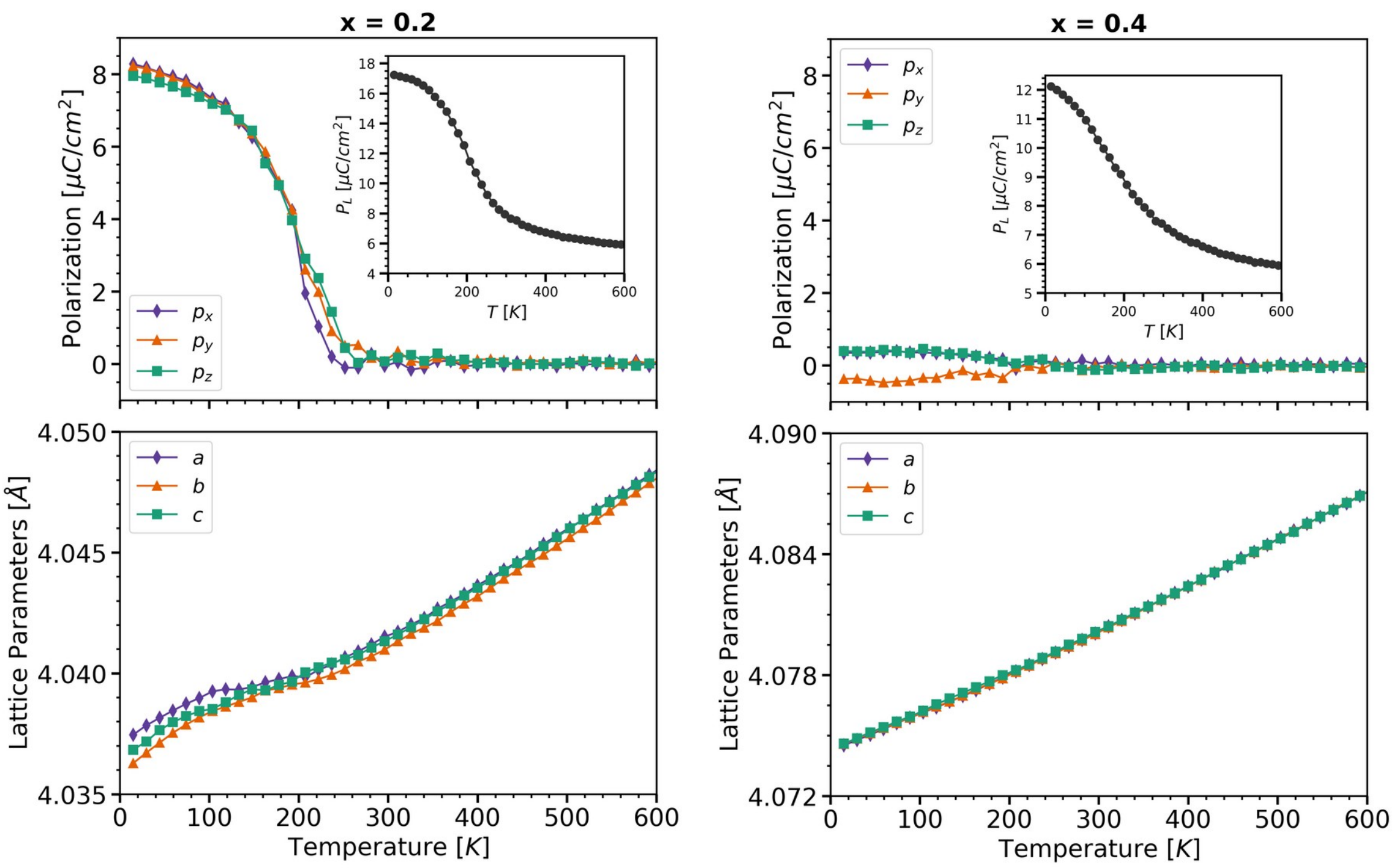

**Figure 1.** Temperature dependence of the macroscopic polarization and lattice parameters for the random solid solution (S0) systems: x = 0.2 (left panel), showing a clear ferroelectric transition, and x = 0.4 (right panel), exhibiting the characteristic behavior of a relaxor where macroscopic polarization is suppressed. The insets display the thermal evolution of the local polarization magnitude $P_L$.

### 3.2. Effect of Zr segregation on structural properties and polarization

The analysis of the averge unit cell volume as a function of temperature reveals that Zr segregation induces a significant lattice expansion compared to the random solid solution (S0), as shown in **Figs. 2a** and **2b** for x=0.2 and x=0.4, respectively.

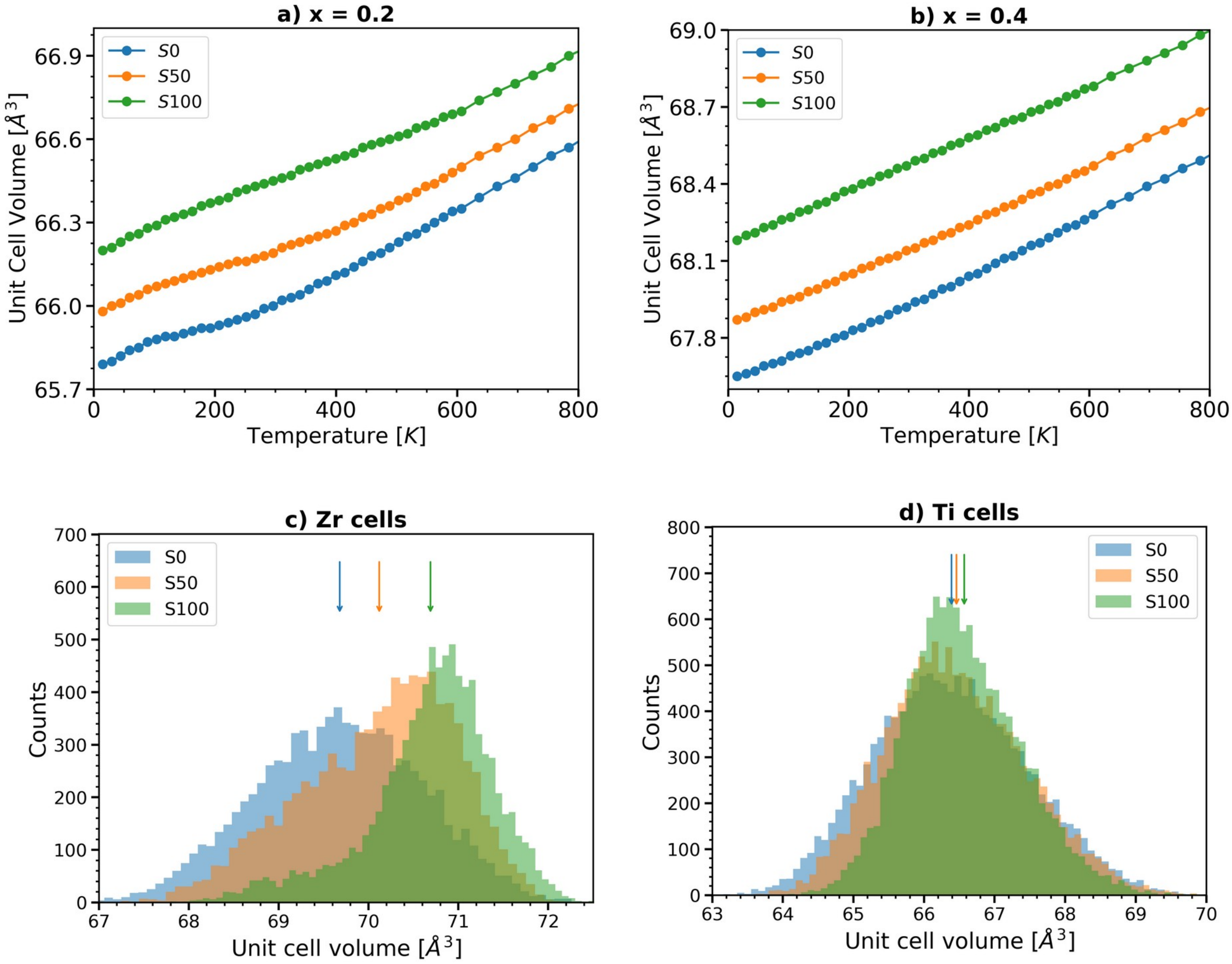

**Figure 2.** Temperature dependence of the average unit cell volume for x=0.2 (a) and x=0.4 (b) at different degrees of Zr segregation (S0, S50, and S100). Distributions of local volume of the perovskite unit cells centered at the Zr-sites (c) and Ti-sites (d) for the x=0.4 composition at T=20 K.

To gain deeper insight into this segregation-induced lattice expansion, we determined the local volume of the perovskite unit cells centered at the B-sites (either Ti or Zr atoms). **Figures 2c** and **2d** present the volume histograms for the x=0.4 composition at T=20 K, with the average values of the distributions indicated by arrows. While the distributions for the Zr-centered cells display a clear shift toward larger volumes as segregation increases (**Fig. 2c**), the effect on the Ti-centered cells is much less pronounced. By analyzing the average values, we can estimate the volume variation within each sublattice. The results indicate that for the 100% segregated system, the volume of the Zr sublattice expands by 1.4% relative to the disordered solid solution, whereas the expansion of the Ti sublattice is only 0.27%. Interestingly, a closer inspection of the Ti distribution reveals that the cells with smaller initial volumes are the most affected, exhibiting the largest expansion upon segregation. Furthermore, the distributions for the 100% segregated system become noticeably narrower, indicating a more uniform volume distribution. This behavior arises because, in the fully segregated state, each sublattice closely resembles its respective bulk material, minimizing local structural fluctuations.

In the random solid solution (S0), the local lattice strains induced by individual, size-mismatched Zr atoms are averaged out and uniformly distributed, effectively dissipating throughout the host matrix without constructive interaction. Conversely, chemical segregation forces these distinct strain fields to accumulate cooperatively. Under these conditions, the $BaZrO_3$-rich nanoregions act as rigid expansion centers that compel the surrounding $BaTiO_3$ matrix to stretch. This localized lattice expansion, concentrated primarily within the Zr-rich inclusions, relieves internal elastic strain in the surrounding Ti-rich matrix, creating an energetically favorable framework that sustains robust Ti off-center displacements.

Polarization behavior in the ferroelectric and relaxor regimes is presented in **Fig. 3**. In the ferroelectric phase (x=0.2), increasing the degree of segregation stabilizes the ordered state, leading to an enhanced macroscopic polarization and a shift of the transition toward higher temperatures (**Fig. 3a**). Conversely, for the relaxor composition (x=0.4), the macroscopic polarization remains zero across the entire temperature range due to the pronounced local disorder (**Fig. 3b**). This suppression of long-range order persists even for the 100% Zr-segregated system, confirming that clustering alone does not restore macroscopic ferroelectricity in the relaxor regime. To probe the microscopic mechanisms governing BZT, we focused our analysis on the local polarization ($P_L$), defined as described in Section 2. This parameter is crucial for characterizing relaxor systems, as it can unveil highly polarized nanoregions even when the net macroscopic polarization averages to zero due to dipolar disorder.

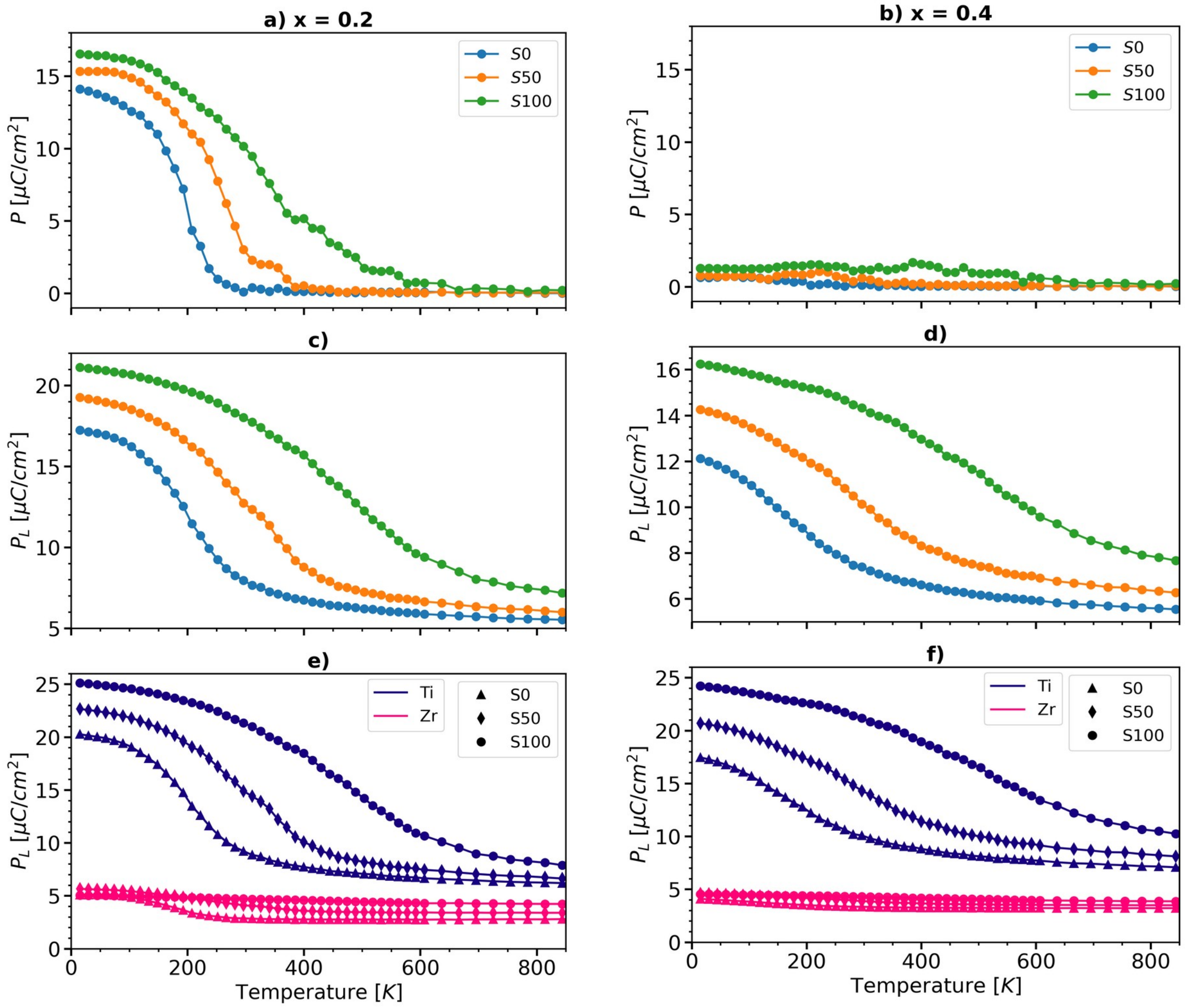

**Figure 3.** Temperature dependence of macroscopic polarizations for x=0.2 (a) and x=0.4 (b) at different degrees of Zr segregation (S0, S50, and S100). Local polarizations as a function of temperature for x=0.2 (c) and x=0.4 (d). (e-f) Local polarization associated with Ti and Zr species as a function of temperature, highlighting the enhancement of Ti-site polarization due to clustering.

The temperature dependence of the local polarization for different degrees of segregation is shown in **Figs. 3c** and **3d**. At low temperatures, all configurations exhibit high local polarization values, regardless of their macroscopic order. Notably, the fully segregated systems (S100) maintain significantly higher local polarization levels than the random solid solution (S0), indicating that Zr clustering effectively stabilizes polar order at the local scale.

A closer look at the species-specific response (**Figs. 3e** and **3f**) reveals that Ti-centered cells exhibit significantly larger polarizations than their Zr-centered counterparts. The latter, however, do not adopt the perfect non-polar cubic bulk structure of $BaZrO_3$, but rather present a small polarization below ~ 5 $\mu C/cm^2$, which does not depend strongly on segregation. The larger dipole moments centered at the Ti sites remain remarkably high even well above the macroscopic phase transition temperature of the ferroelectric phase. This persistent local polarization at high temperatures originates from the lattice distortions induced by the size mismatch of the Zr atoms, which break the local cubic symmetry and sustain Ti off-center displacements within the macroscopically paraelectric regime. Importantly, the polarization enhancement upon segregation cannot be explained solely by volume expansion in the Ti sublattice, since this structural change is remarkably small (0.27%), and no correlation was found between local polarization and unit cell volume, as shown in Fig. S1 of the Supplemental Material. The underlying mechanism driving this behavior is discussed in detail in section 3.4.

### 3.3. Dielectric Response and Phase Transitions

To evaluate how this enhanced local polarization influences the macroscopic dielectric response, we determined the temperature dependence of the relative dielectric permittivity ($\varepsilon$). The permittivity was calculated through a direct method by applying a small perturbing external electric field along the [111] direction during MD heating simulations. The resulting polarization was then used to evaluate the permittivity components. Figure 4 shows $\varepsilon$ as a function of temperature for both the ferroelectric ($x = 0.2$) and relaxor ($x = 0.4$) compositions across different degrees of chemical segregation. For the $x = 0.2$ system (**Fig. 4a**), the permittivity exhibits a peak at the Curie temperature, typical of a standard ferroelectric transition where long-range order is established. The response for the x = 0.4 composition (**Fig. 4b**) is characterized by a flattened permittivity maximum at $T_m$. This diffuse profile is a fundamental signature of the relaxor state, directly linked to the absence of macroscopic polarization at zero field. In this regime, the system does not undergo a global symmetry-breaking transition.

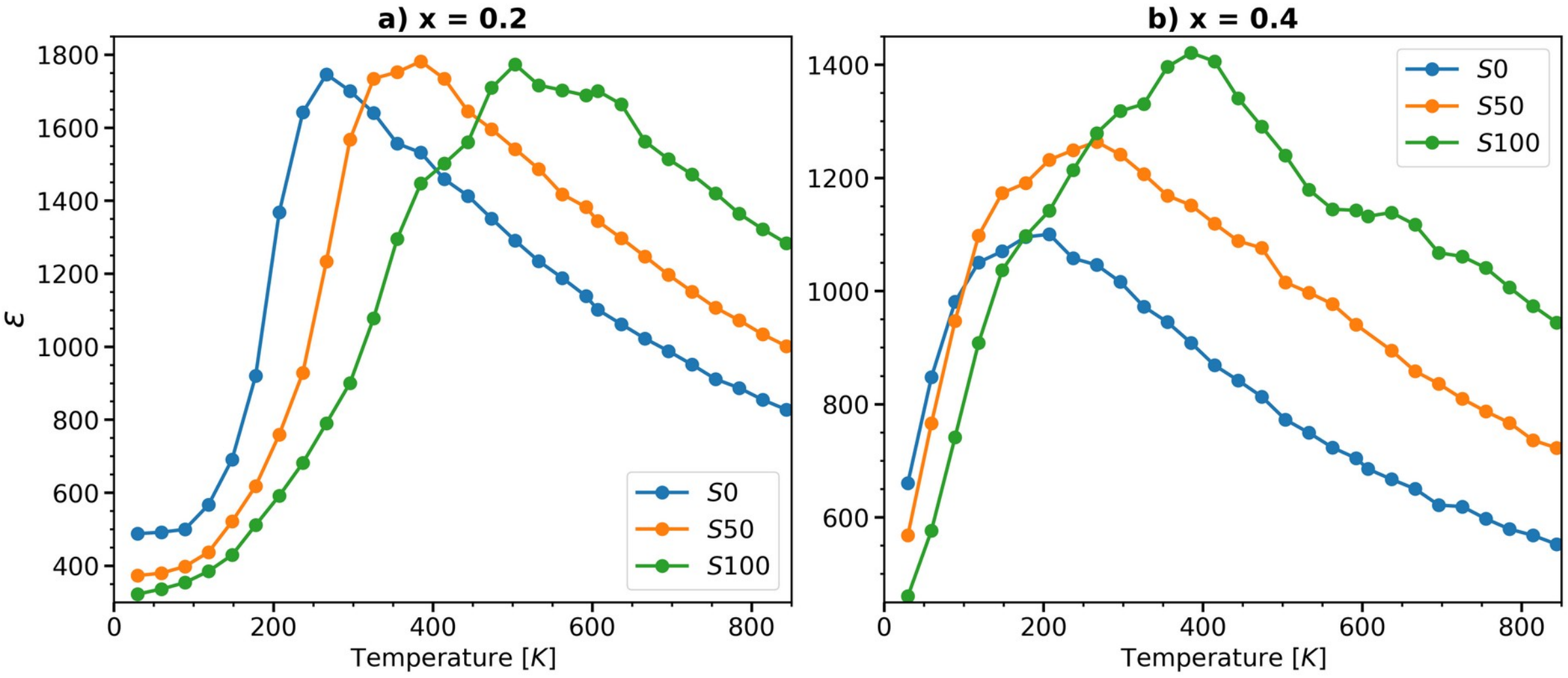

**Figure 4.** Temperature dependence of the relative dielectric permittivity for (a) x=0.2 and (b) x=0.4 at different degrees of Zr segregation (S0, S50, and S100). The broadened peak in x=0.4 illustrates the relaxor nature of the phase, where the lack of a sharp transition corresponds to the suppression of macroscopic polarization.

Instead, the diffuse dielectric response reflects the progressive freezing of local polar regions (such as polar nanoregions or slush-like structures) that remain decoupled, preventing the emergence of a net macroscopic polarization even at the lowest temperatures. As shown in **Fig. 4b**, Zr segregation strongly modulates this behavior, reinforcing the local nature of the polar states without inducing a long-range ferroelectric phase.

Notably, the impact of chemical clustering reveals a distinct behavior between the two regimes. For the ferroelectric composition, Zr segregation systematically shifts the permittivity peak toward higher temperatures without significantly altering its maximum value. Conversely, for the relaxor regime, chemical segregation not only shifts $T_m$ to higher temperatures but also drives a substantial enhancement in the peak permittivity. This dual effect underscores that in the relaxor phase, the cooperative accumulation of local elastic strains not only provides thermal stability to the polar structures but also enhances their dielectric susceptibility under external fields, ultimately maximizing the material's functional performance.

### 3.4. Local Structural Analysis and Chemical Environment

The macroscopic polar response of BZT is deeply rooted in the distribution of local structural phases. By employing a machine learning classification approach, we quantified the fraction of Ti-centered cells belonging to the cubic (C), tetragonal (T), orthorhombic (O), and rhombohedral (R) phases as a function of temperature. Zr cells are excluded from this analysis due to their intrinsically nonpolar nature, which results in an ambiguous classification. Further details regarding this procedure can be found in Ref. [14]. The results for the ferroelectric ($x = 0.2$) and relaxor ($x = 0.4$) compositions are shown in **Figs. 5** and **6**, respectively. A key observation from these figures is that the system does not exhibit a single, uniform local symmetry. Instead, a coexistence of R, O, and T phases occurs at low temperatures. This structural heterogeneity is directly linked to the specific chemical environment of each Ti cation [14,25].

As demonstrated by the titanium neighbor histograms, Zr segregation significantly alters the local coordination. In the S100 system, the population of Ti cells surrounded by six Ti neighbors increases markedly. According to the conditional probability framework, Ti cells with six Ti neighbors are more likely to adopt the rhombohedral symmetry [14,25]. Consequently, Zr clustering effectively favors the R ferroelectric configuration, shifting the onset of the cubic paraelectric phase toward much higher temperatures compared to the random solid solution (S0). From a mesoscopic perspective, this segregation creates larger, continuous Ti-rich spatial regions that accommodate extended domains of rhombohedral symmetry.

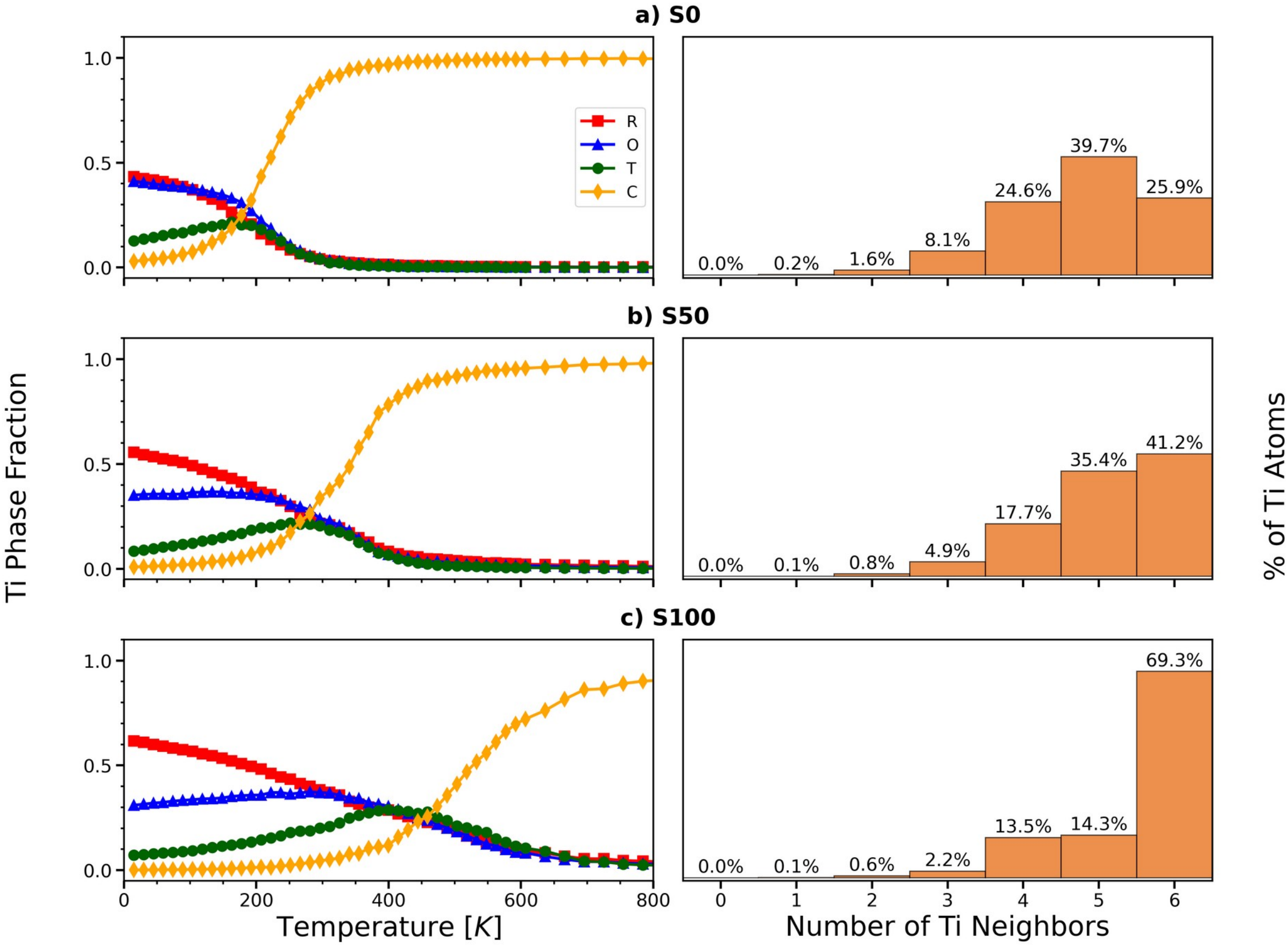

**Figure 5.** Fraction of rhombohedral (R), ortorhombic (O), tetragonal (T), and cubic (C) Ti-centered cells in the ferroelectric phase of BZT (x = 0.2) as a function of temperature, alongside Ti-neighbor histograms for each case: **(a)** S0, **(b)** S50, and **(c)** S100. The plots illustrate how Zr segregation favors the stabilization of local ferroelectric phases by modifying the Ti-Ti coordination environment.

By minimizing the structural frustration typically induced by random Zr-cell intrusion, these extended regions facilitate a strong cooperative alignment among neighboring dipoles, thereby significantly increasing the dipolar correlation length. This reduction in local structural frustration promotes longer-range dipolar interactions, providing a clear microscopic mechanism for the enhanced spatial correlations and lower phase dispersion previously identified in Ref. [14]. This stabilization and enhanced spatial correlation are particularly significant because the local polarization associated with the R phase is intrinsically higher than that of the O phase, which in turn exceeds that of the T phase, following the hierarchy $P_L(R) > P_L(O) > P_L(T)$.

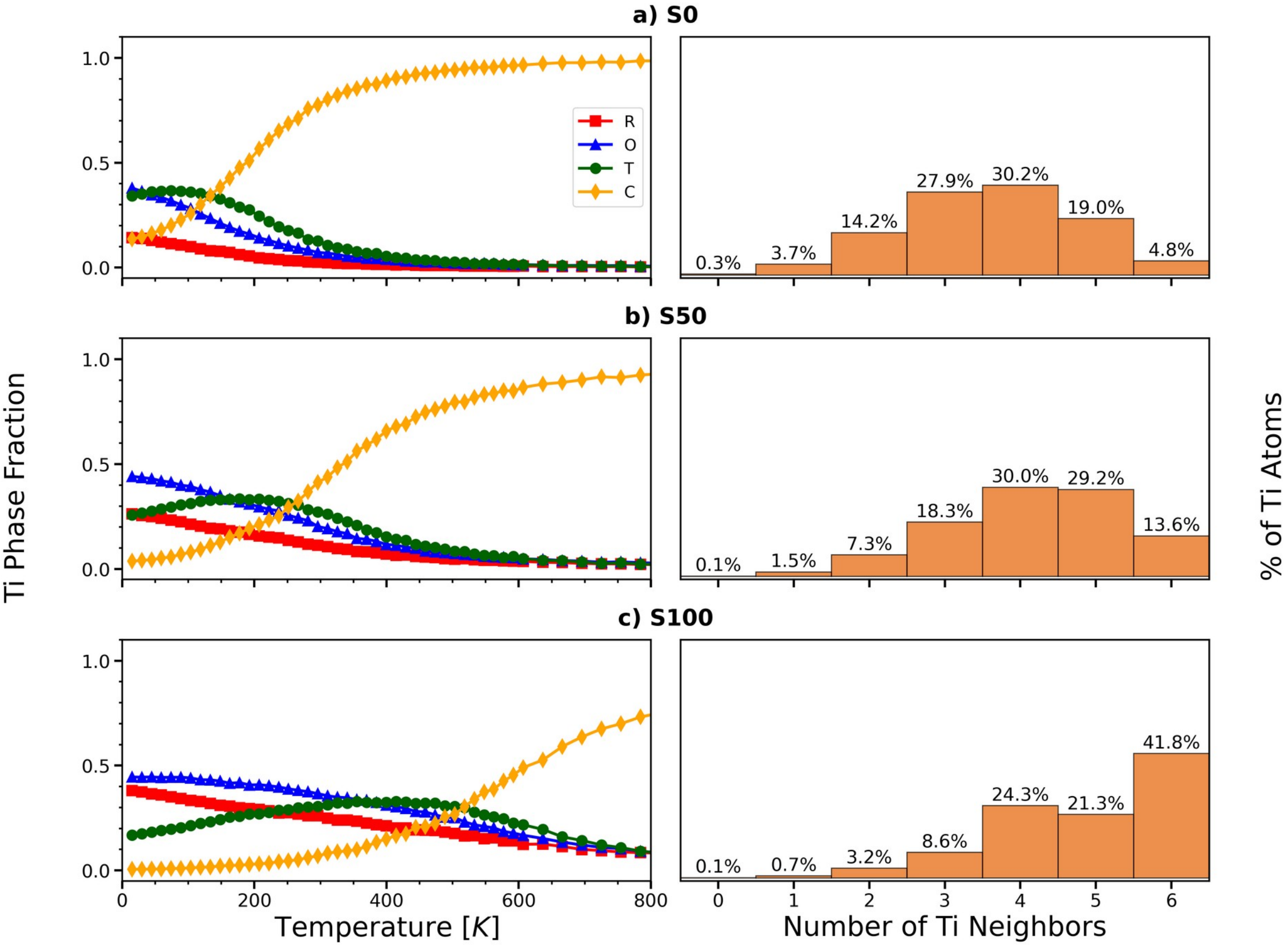

**Figure 6.** Fraction of rhombohedral (R), ortorhombic (O), tetragonal (T), and cubic (C) Ti-centered cells in the relaxor phase of BZT (x = 0.4) as a function of temperature, along with Ti-neighbor histograms for each segregation level: **(a)** S0, **(b)** S50, and **(c)** S100. The data highlights the persistence of local polar phases at high temperatures induced by Zr clustering.

Detailed values for the x = 0.4 composition at T= 20K are provided in Fig. S2 and Table S1 of the Supplemental Material. This polarization hierarchy among the local phases mirrors the behavior observed for the different ferroelectric phases of bulk $BaTiO_3$ and holds true across all studied segregation degrees. Ultimately, this cooperative stabilization of highly polar local phases and their extended spatial correlations provide a clear microscopic explanation for the enhanced polarization observed in the segregated systems (Figs. 3e and 3f).

A significant finding in the x = 0.4 systems is the persistence of local polar phases even at temperatures far above the permittivity maximum $T_m$. While the fraction of cubic (C) cells increases with temperature, Zr segregation effectively preserves a non-negligible population of polar cells in the high-temperature regime. In the case of maximum segregation (S100), the local Ti-rich environments remain highly polarized even in the absence of macroscopic order. This result suggests that Zr clustering reinforces the stability of local polar clusters.

### 3.5. Determination of Burns and Deviation temperatures

The persistence of the local polar phases identified previously is closely related to the nucleation of polar nanoregions at a characteristic thermal threshold. In relaxor systems, this threshold is defined as the Burns temperature ($T_B$), which marks the point during cooling where local dipoles start to correlate, even though the macroscopic symmetry remains cubic. For the $BaZr_{0.4}Ti_{0.6}O_3$ composition (x = 0.4), $T_B$ was estimated to be around 440 K from dilatometry experiments [4, 31]. Interestingly, the $T_B$ value obtained from the inverse dielectric permittivity curve is much lower than that derived from the thermal strain behavior [31, 32].

To provide a robust characterization of $T_B$ in the x=0.4 relaxor phase, we implemented three independent numerical methods based on different physical properties of the system. The results of these estimations are presented in **Fig. 7**. The first method (a) identifies $T_B$ by detecting the deviation of the dielectric permittivity from the Curie-Weiss law at high temperatures. The second approach (b) analyzes the thermal evolution of the unit cell volume, identifying $T_B$ as the temperature where the lattice expansion deviates from the linear regime —an analysis equivalent to experimental dilatometry assays [4,31]. Finally, the third method (c) monitors the temperature dependence of the local polarization magnitude, establishing a criterion based on tangent lines to pinpoint the onset of significant local polar order. As shown across the columns of **Fig. 7**, the consistent estimation of the Burns temperature ($T_B$) through three independent methods confirms that $T_B$ systematically increases with the degree of Zr segregation.

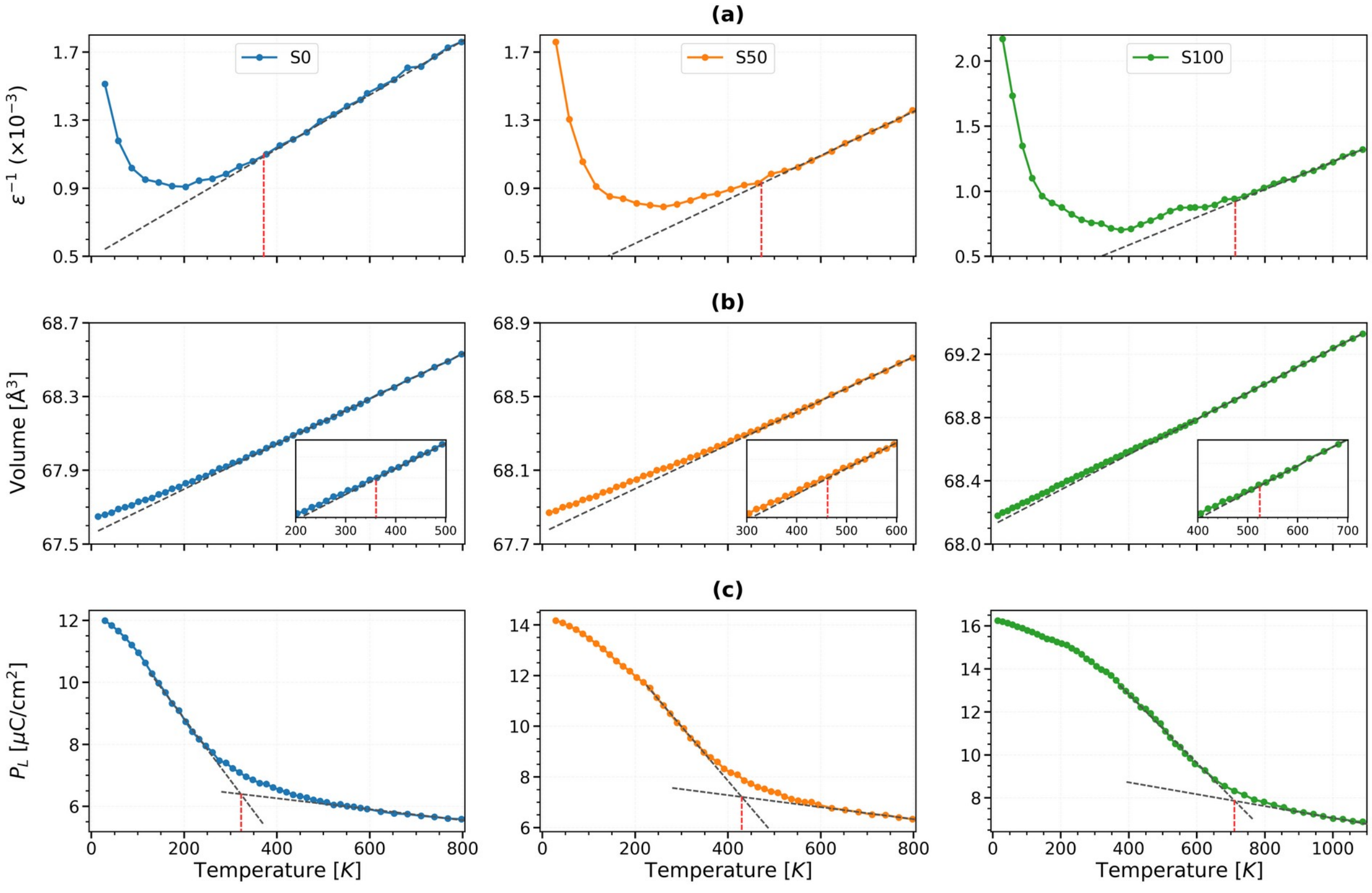

**Figure 7.** Graphical representation of the three independent methods used to determine the Burns temperature (TB) for the BZT relaxor phase (x=0.4) at different segregation levels: (a) Deviation from the Curie-Weiss law in dielectric permittivity, (b) non-linear deviation in unit cell volume, and (c) thermal evolution of local polarization. Each column represents a different degree of segregation (S0, S50, and S100), with red dashed lines indicating the estimated TB values

For the S0 configuration, the $T_B$ values obtained from these distinct methods are closely grouped at 380, 360, and 330 K, and they systematically shift upward to 470, 460, and 430 K for the S50 system. Consequently, we can estimate $T_B = 360 \pm 30$ K for S0 and $450 \pm 20$ K for S50, while a larger uncertainty is found for the S100 system, yielding $640 \pm 90$ K. This upward shift demonstrates that chemical clustering not only increases the local polarizations at lower temperatures but also promotes an earlier nucleation of polar nanostructures during the cooling process. Crucially, the temperature interval between $T_B$ and the permittivity maximum $T_m$ widens with increasing segregation. This trend strongly suggests that chemically segregated systems exhibit a significantly broader temperature window where "slush-like" dynamics or polar nanoregion (PNR) fluctuations dominate the material's overall response.

To maintain a consistent analysis across both compositions, we applied the same three diagnostic methods to the diffuse ferroelectric phase (x = 0.2) to identify a deviation temperature ($T_D$). While the Burns temperature $T_B$ is evaluated for the relaxor composition (x = 0.4) to mark the onset of dynamic polar nanoregions, $T_D$ captures the crossover in the diffuse ferroelectric regime where local dipoles develop significant correlations and the system deviates from high-temperature paraelectric behavior. This behavior is consistent with experimental observations; for instance, the phase transition behavior of $BaZr_{0.2}Ti_{0.8}O_3$ in the paraelectric region indicates the presence of local structures well above the paraelectric-ferroelectric transition temperature ($T_C$ = 292 K) [33]. Specifically, a non-zero remnant polarization was experimentally measured up to a characteristic temperature of 350 K, which closely coincides with the temperature where the dielectric constant deviates from the Curie–Weiss law [33].

Building upon this experimental baseline, the results for the x = 0.2 systems are summarized in **Fig. 8**. Similar to the relaxor observations, $T_D$ can be consistently identified using the three previously established methods. In all cases, $T_D$ is found to be systematically higher than the macroscopic Curie temperature —yielding 370 ± 90 K for S0, 480 ± 60 K for S50, and 600 ± 50 K for S100 — typically by approximately 100 K. As observed in Fig. 8a-c, Zr segregation also shifts $T_D$ toward higher temperatures in the ferroelectric phase. However, the difference between $T_D$ and the Curie temperature remains relatively constant across different segregation levels in this regime. This contrast with the relaxor phase—where the gap between $T_B$ and $T_m$ increases with clustering—highlights that while segregation stabilizes local order in both cases, its impact on the diffusivity of the transition is specifically more pronounced in the relaxor composition.

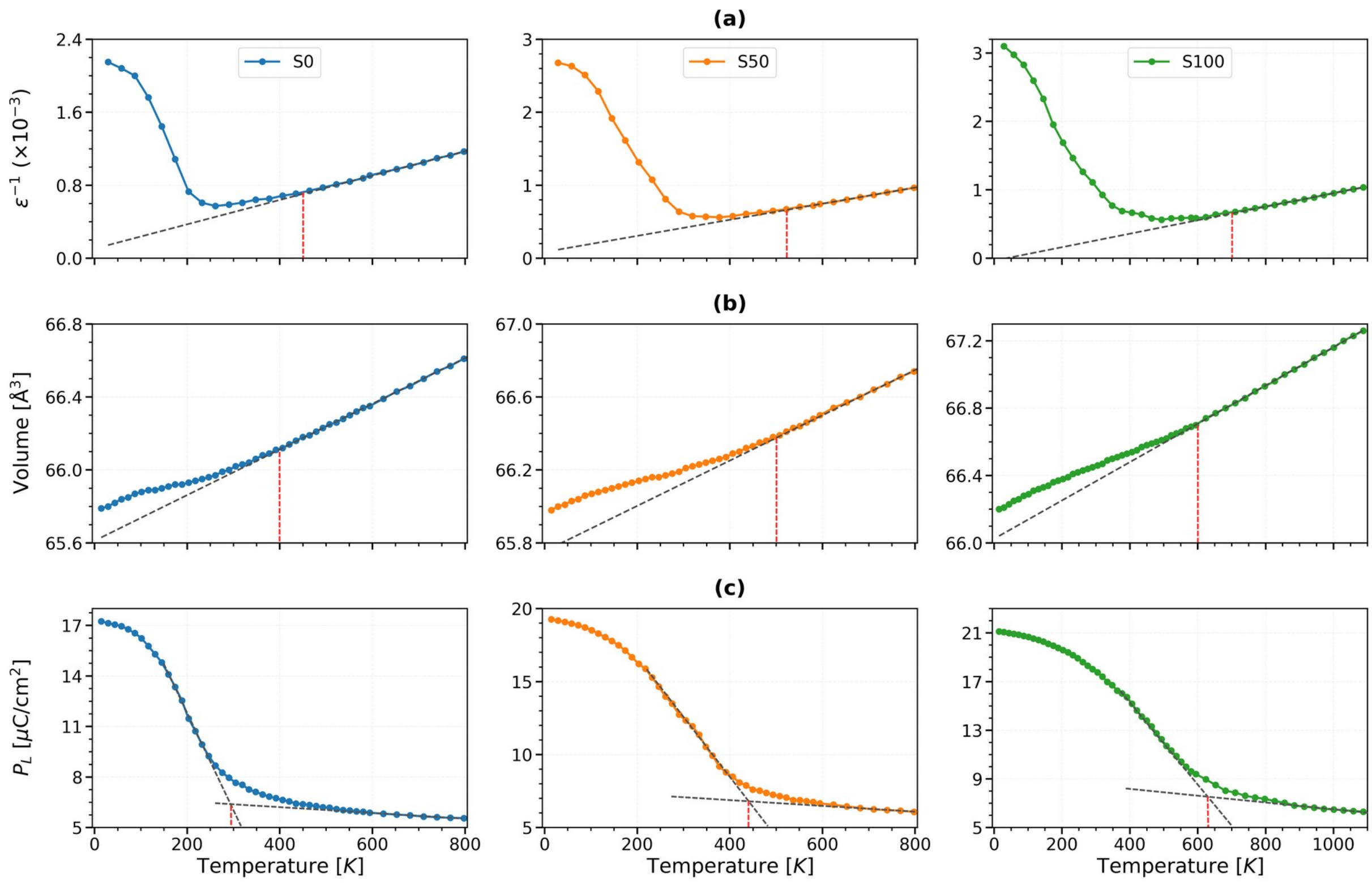

**Figure 8.** Graphical representation of the methods used to determine the deviation temperature ($T_D$) for the BZT ferroelectric phase (x = 0.2) at various segregation levels : (a) Deviation from the Curie-Weiss law in dielectric permittivity, (b) non-linear deviation in unit cell volume, and (c) thermal evolution of local polarization. Each column represents a different degree of segregation (S0, S50, and S100), with red dashed lines indicating the estimated $T_D$ values.

### 3.6. Ferroelectric hysteresis loops and polarization retention

The ferroelectric nature and switching dynamics of the BZT systems were evaluated via P-E hysteresis loops at 100 K, as shown in **Fig. 9**. The results highlight a profound impact of chemical segregation on the macroscopic ferroelectric response for both compositions. For the ferroelectric regime (x = 0.2, **Fig. 9a**), the random solid solution (S0) displays a standard hysteresis loop. However, with increasing Zr segregation, both the remnant polarization ($P_r$) and coercive field ($E_c$) increase, most notably $E_c$. This electrical 'hardening' suggests that Zr clustering stabilizes local polar domains, rendering them more resilient to field-driven reorientation.

In the relaxor regime (x = 0.4, **Fig. 9b**), the random solid solution (S0) exhibits a characteristic slim hysteresis loop with negligible remnant polarization. Interestingly, chemical segregation induces a similar hardening trend, progressively elevating $P_r$ and driving a pronounced rise in $E_c$. This demonstrates that chemical clustering restricts polarization switching across both structural regimes.

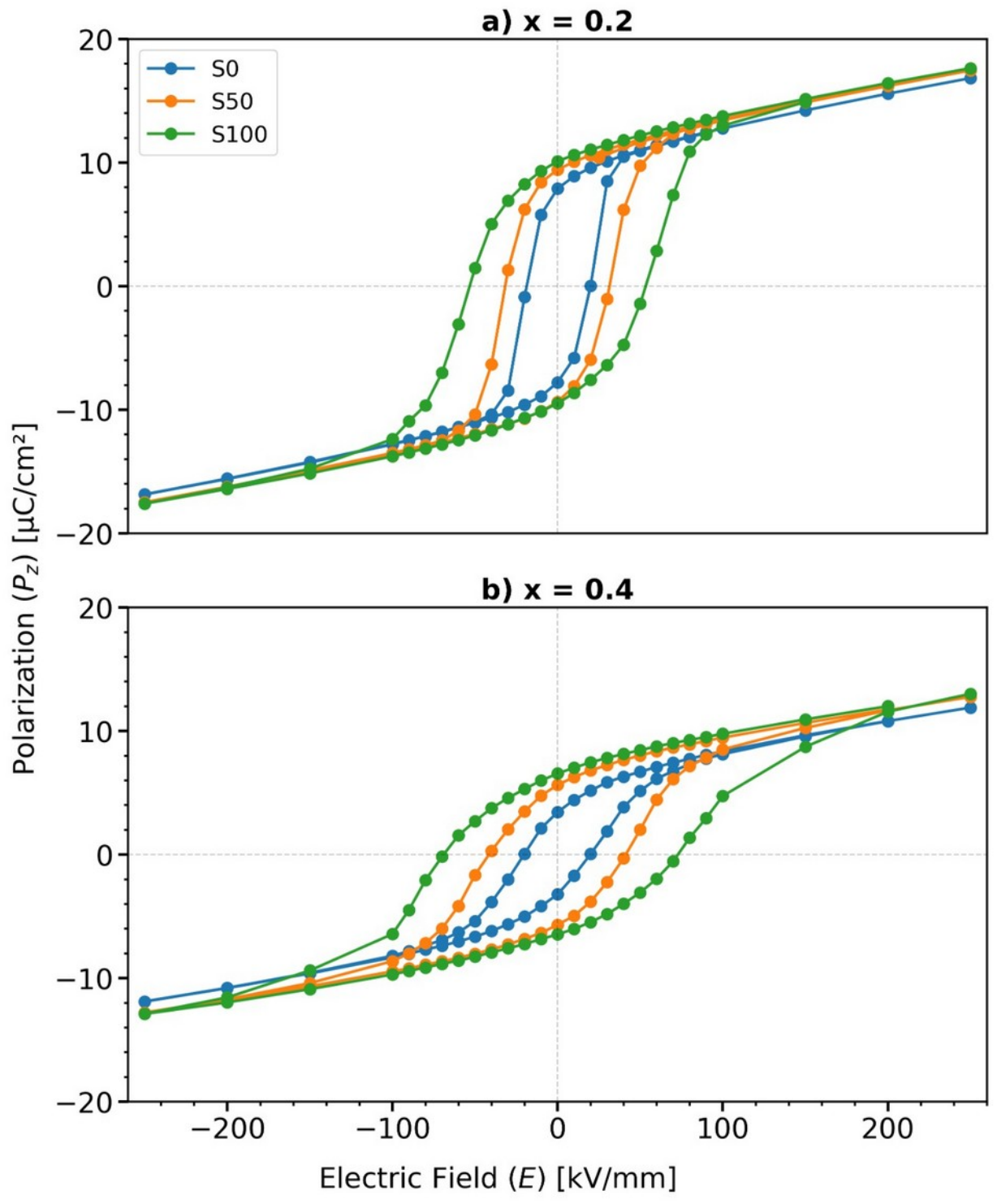

**Figure 9.** Polarization-Electric field (P-E) hysteresis loops for the different degrees of Zr segregation: **(a)** Ferroelectric phase (x = 0.2) and **(b)** relaxor phase (x = 0.4). The plots illustrate the increase in coercive field and polarization magnitude induced by chemical clustering.

To further elucidate the dynamic stability of the field-aligned states in the relaxor phase, we evaluated the temporal polarization retention for the x = 0.4 configurations at various temperatures below $T_m$. **Figure 10a** shows the decay of the remnant polarization as a function of time following the removal of the external electric field for the three Zr configurations at 150 K. The results reveal a clear correlation between chemical segregation and the long-term stabilization of polar states; specifically, the S0 configuration exhibits a significantly faster polarization decay than its segregated counterparts.

The relaxation dynamics were fitted using a stretched exponential function (Kohlrausch-Williams-Watts law) $P(t) / P_0 = \exp[- (t/\tau)^{\beta}]$, widely employed in relaxor ferroelectrics [34,35]). At 150 K, the resulting fit parameters are $\tau$ = 145±3, 7700±600, and 77000±7000 ps,

with stretching exponents of β = 0.485±0.008, 0.43±0.01, and 0.371±0.006 for the S0, S50, and S100 configurations, respectively. The systematic decrease in the stretching exponent β reflects a progressive broadening in the distribution of relaxation times, driven by the strong spatial energy landscape fluctuations introduced by the segregated nanoclusters. The corresponding stretched exponential fits are depicted as dashed lines.

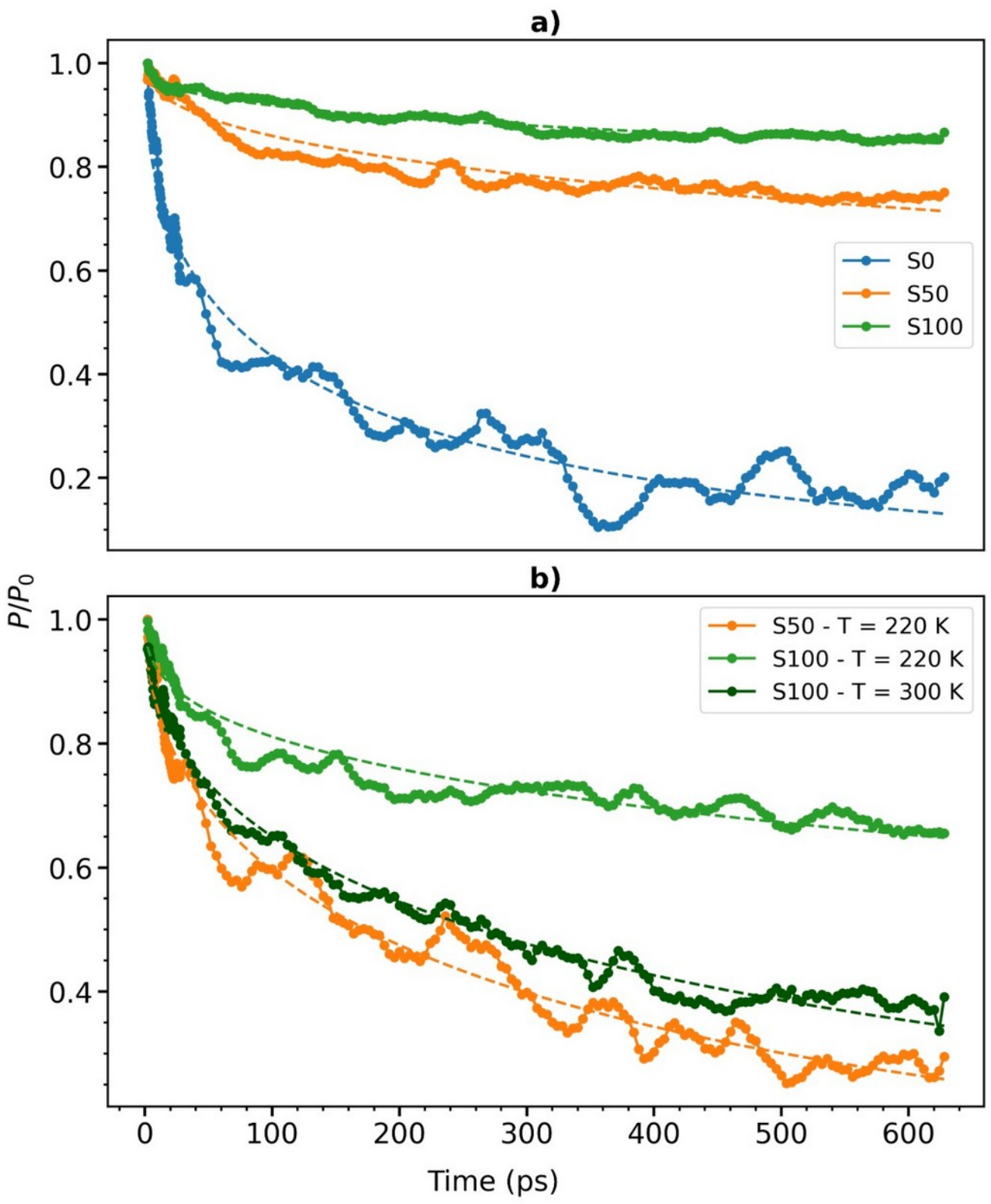

**Figure 10.** Temporal retention of remnant polarization for BZT (x = 0.4) under different chemical segregation states (S0, S50, and S100) after the removal of the external electric field. (a) Normalized polarization decay P(t)/P0 as a function of time at 150 K. Dashed lines represent non-linear fits using the stretched exponential (Kohlrausch-Williams-Watts) model. (b) Polarization retention dynamics for the segregated configurations (S50 and S100) at higher temperatures (220 K and 300 K), illustrating the temperature dependence of the relaxation time τ.

Notably, while S50 and S100 maintain robust retention at 150 K (plateauing above 75% and 85% of $P_0$, respectively), increasing the temperature systematically drives the decay toward significantly lower remnant polarization levels (**Fig. 10b**). At elevated temperatures (220 K and 300 K), thermal fluctuations provide sufficient energy to partially overcome the pinning barriers imposed by the Zr clusters, resulting in lower asymptotic plateaus. Nevertheless, complete chemical segregation continues to offer substantial thermal resistance: the retention profile of S100 at 300 K closely matches that of S50 at 220 K, demonstrating that full Zr segregation compensates for thermal depolarization by an equivalent shift of approximately 80 K.

Quantitative analysis of this temperature dependence via the stretched exponential fits reveals a sharp decline in the relaxation time with increasing temperature. At T = 220 K, the fitted parameters yield $\tau$ = 351±5 ps ($\beta$ = 0.521±0.007) for S50 and $\tau$ = 5100±300 ps ($\beta$ = 0.396±0.009) for S100, whereas at T = 300 K, the S100 system yields $\tau$ = 552±6 ps ($\beta$ = 0.492±0.005). These results confirm that despite the enhanced domain stability and elevated $P_r$, the segregated configurations retain their intrinsic relaxor dynamics—even in the fully segregated limit (S100)—as evidenced by the broad distribution of relaxation times ($\beta < 0.5$) and the strongly temperature-activated decay of $\tau$.

**3.7. Emergence of swirling topological states in the relaxor regime**

Beyond modifying the macroscopic electrical response, chemical segregation induces a profound spatial reorganization within the polarization field texture. Specifically, the localized strain fields surrounding the Zr-rich clusters break the traditional domain-wall paradigms of standard ferroelectrics, promoting the formation of non-trivial, complex topological configurations.

To elucidate these spatial textures in the relaxor phase (x = 0.4), we first analyze the spatial distributions of two key local field invariants: the divergence and the curl (rotational) of the polarization field. **Figure 11** displays the probability distributions for both the divergence (left) and the magnitude of the curl (right) of P across the three Zr configurations at low temperature (15 K). For clarity, the contributions from Ti-centered and Zr-centered unit cells are resolved separately. Notably, the divergence distribution remains largely insensitive to the degree of Zr segregation. In particular, the mean values are virtually identical across all three cases (-1.5 x $10^8$ μC/cm$^3$ for Zr-centered cells and 1.0 x $10^8$ μC/cm$^3$ for Ti-centered cells). The distributions reveal that Zr-centered unit cells exhibit almost exclusively negative divergence values, a characteristic feature of an anti-hedgehog polar arrangement [29]. To verify physical consistency, the volume integral of the polarization divergence over the whole supercell vanishes within computational tolerance for all evaluated configurations.

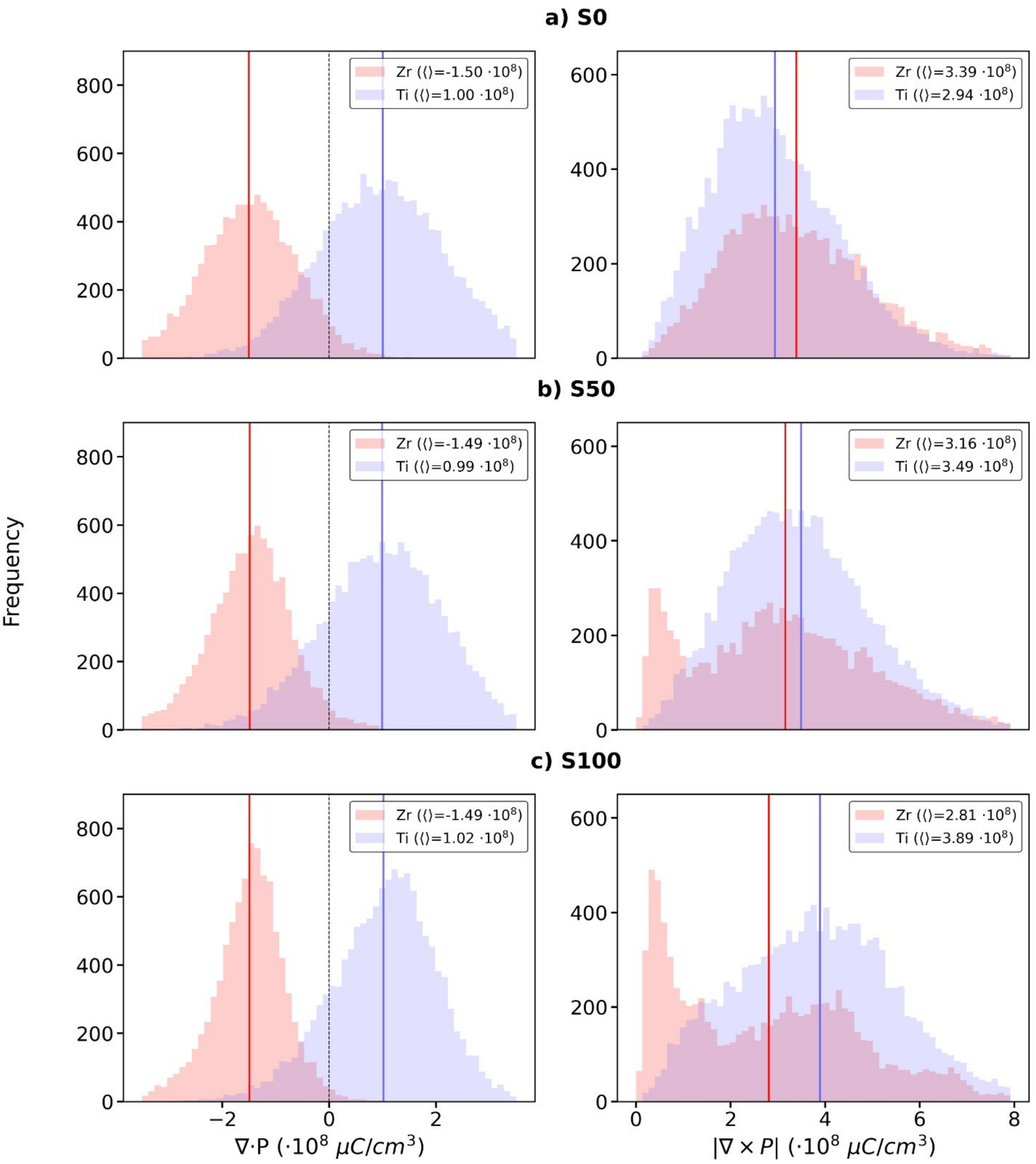

**Figure 11.** Spatial distribution of local polarization field invariants for the S0 (a), S50 (b), and S100 (c) Zr configurations in BZT (x = 0.4) at 15 K, resolved separately for Ti-centered and Zr-centered cells. (left) Probability distributions of the divergence of P, highlighting the negative divergence character of Zr-centered unit cells indicative of anti-hedgehog configurations. (right) Probability distributions of the magnitude of the curl of P, illustrating the enhancement of vortex-like swirling textures in Ti-rich regions upon Zr segregation. Insets display the corresponding mean values expressed in μC/cm$^3$.

In contrast, the distribution of the magnitude of curl P is strongly modulated by chemical segregation (Fig. 11, right column). For Zr-centered cells, the mean curl magnitude systematically decreases with increasing segregation, as evidenced by the emergence of a pronounced peak near zero in S50 and S100. This indicates that the core of the Zr-rich clusters becomes topologically uniform with vanishing vorticity. Conversely, for Ti-centered unit cells, the mean curl magnitude increases from $2.94 \times 10^8$ $\mu C/cm^3$ in S0 to $3.89 \times 10^8$ $\mu C/cm^3$ in S100. This shift reflects a marked enhancement in dipolar curling and indicates that, if vortex-like textures are present, their rotational activity is preferentially localized within the surrounding Ti-rich matrix. Accordingly, we identify the presence of these vortex-like structures by combining direct spatial visualization of local polarization vectors with mapping the magnitude of the polarization curl, where regions exhibiting pronounced local values serve as possible locations for vortex cores.

This spatial redistribution is visually confirmed in **Fig. 12**, which compares two-dimensional cuts of the polarization vector field for S0 and S100 at 15 K (on cooling), color-coded by the out-of-plane curl component to highlight local vorticity. While S0 exhibits spatially fragmented and low-intensity curl fluctuations, S100 develops well-defined, highly coherent vortex cores with opposite senses of rotation (indicated by intense blue and red regions representing clockwise and counter-clockwise swirling, respectively).

Crucially, these high-vorticity cores are embedded within the Ti-rich matrix, while the compact Zr clusters serve as physical boundaries that accommodate the swirling polar flux without suppressing the overall dipole coupling. As clearly observed, chemical segregation significantly promotes the formation of vortex-like structures, which can be categorized into two distinct spatial regimes: vortices circulating around Zr-segregated clusters and vortices localized within the interstitial regions bounded by these clusters. These two vortex configurations closely resemble the topological arrangements previously identified in $BaTiO_3$–$BaZrO_3$ nanocomposites forming vortex supercrystal states [29].

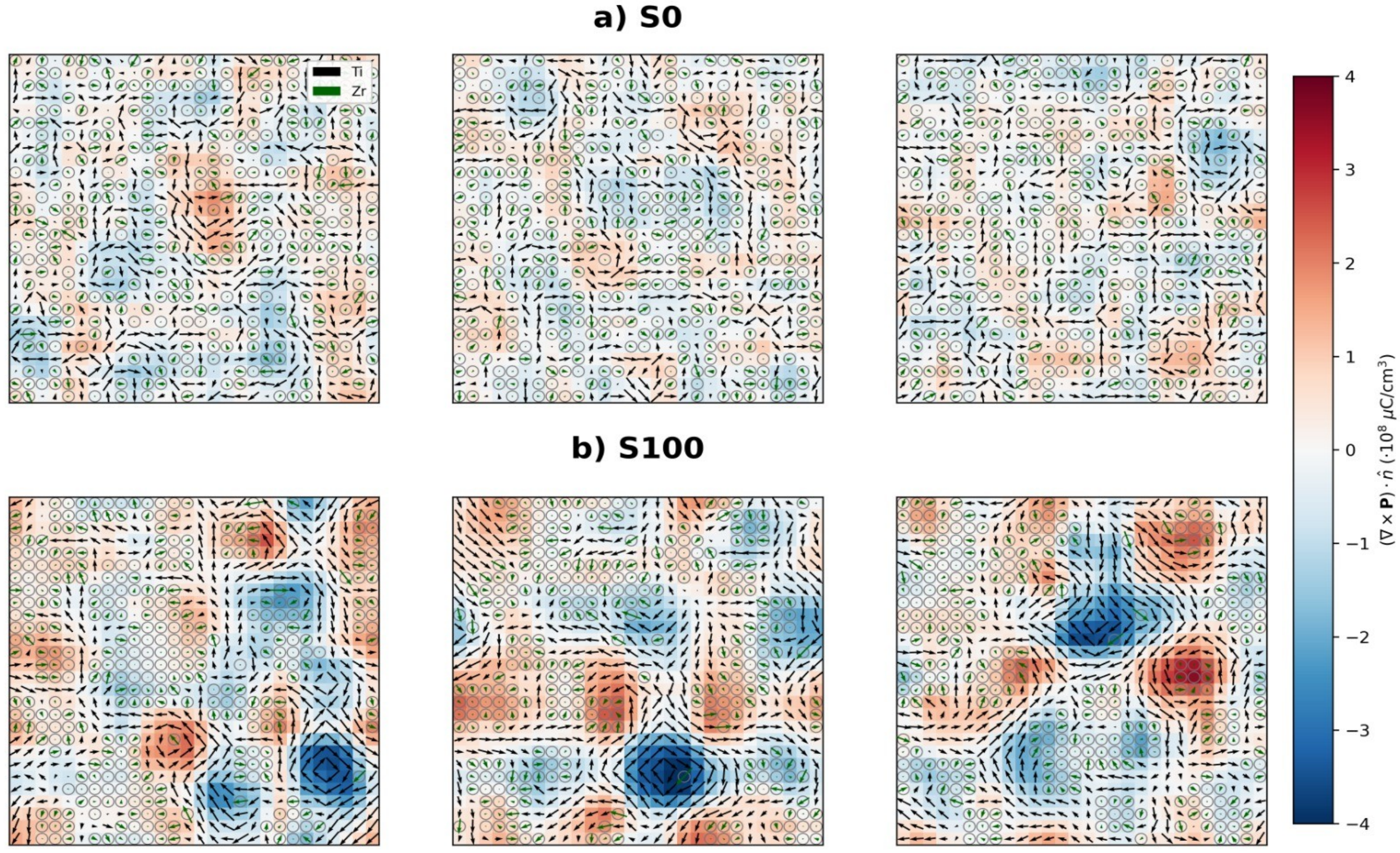

**Figure 12.** Cross-sectional polarization field maps for S0 (a) and S100 (b) configurations in BZT (x = 0.4) at 15 K across three planes. Color regions indicate the out-of-plane component of the curl of P, and circles mark Zr-centered cells. Zr segregation promotes two distinct vortex regimes: cluster-centered and interstitial vortices. Additional slices are shown in Figs. S3 and S4 of the Supplemental Material.

Three-dimensional representations of these two distinct swirling patterns are shown in **Figs. 13a and 13b**. By organizing into these flux-closure vortex configurations, the system effectively accommodates local polar order while minimizing bound charge accumulation, thereby mitigating depolarizing fields at the cluster interfaces and stabilizing these nanoscale polar textures. These two types of vortex-like structures display remarkable thermodynamic stability, persisting across a broad temperature range. This indicates that their emergence is intrinsically linked to the chemical heterogeneity of the host lattice (polarization profiles for the S100 configuration at room temperature are available in Fig. S5 of the Supplemental Material).

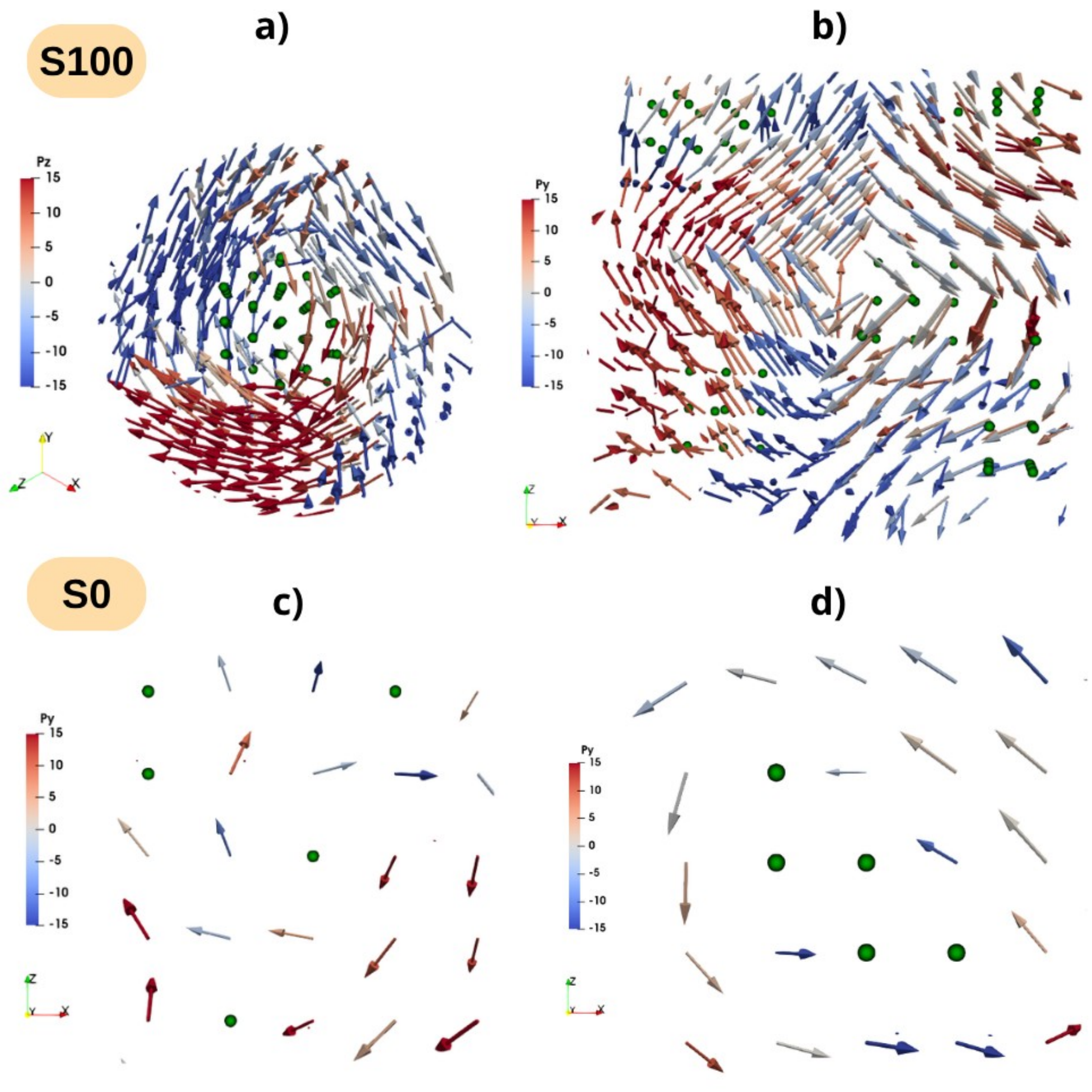

**Figure 13.** Spatial distribution of polar vortex textures in BZT configurations. (a, b) Three-dimensional views of resilient vortices in the fully segregated S100 matrix, showing (a) cluster-encircling and (b) interstitial arrangements. (c, d) Incipient polar vortices in the random S0 alloy nucleated by (c) single Zr substitutions and (d) small stochastic Zr clusters. Green spheres indicate Zr atoms within the Ti-rich host lattice, highlighting how local compositional heterogeneity drives polarization rotation from single-atom to mesoscopic scales. Polarizations are expressed in $\mu C/cm^3$ .

To gain deeper insight into the internal structure of these topological objects, we evaluated the local phase composition within both the cluster-encircling and interstitial vortex regions as a function of temperature (detailed in Fig. S6 of the Supplemental Material). This quantitative analysis reveals that the core and boundaries of both vortex types are predominantly governed by a cooperative coexistence of rhombohedral and orthorhombic local motifs at low temperatures, which smoothly transition into the high-temperature cubic phase. This demonstrates that these swirling patterns are stabilized by the same local structural matrix that underpins the enhanced cooperative polarization of the system (see Fig. 6c). Consequently, their robust presence offers a clear microscopic rationale for the enhanced coercive field observed in our hysteresis loop analysis. Overcoming the energy barrier for macroscopic polarization switching requires the destabilization and unpinning of these resilient topological states, thereby electrically hardening the material.

Interestingly, the formation of polar vortices is not restricted to the fully segregated regime. A closer examination of the stochastically distributed S0 configuration (**Fig. 12a**) reveals incipient, highly localized vortex structures embedded within the random matrix. These confined swirling patterns typically organize around isolated Zr ions (**Fig. 13c**), Zr pairs or small, stochastically formed Zr regions (**Fig. 13d**) that naturally occur in the random alloy. Although less frequent and topologically more transient than those in S100, these S0 vortices demonstrate that even short-range compositional fluctuations are capable of breaking local spatial symmetry and initiating polarization rotation. This indicates that the core mechanism driving vortex formation—namely, the off-centering mismatch between Zr and Ti environments—is intrinsically active at the single-atom level, with macroscopic chemical segregation serving to amplify, align, and globally stabilize these topological textures.

Looking forward, the identification of these resilient topological configurations opens new avenues for investigating the broader mesoscopic nature of relaxor ferroelectrics. Because these closed, vortex-like arrangements inherently confine polarization at the nanoscale, future work will focus on quantifying how different degrees of chemical segregation alter the spatial correlation length and relaxation dynamics of polar displacements. Specifically, we aim to evaluate how the clustering of Zr ions modulates the spatial boundaries and topological connectivity of the polar phase, shedding light on whether chemical segregation favors the classical framework of isolated polar nanoregions or stabilizes a more continuous, interconnected "slush-like" polar network. Understanding how local chemical heterogeneity modulates the boundary between these two competing relaxation paradigms will be crucial for achieving deterministic control over nanoscale polar fields and optimizing the functional performance of lead-free perovskites.

## 4. Conclusions

This study provides a comprehensive atomistic framework detailing the influence of chemical segregation on the structural, functional, and topological polar states of lead-free $BaZr_xTi_{1-x}O_3$. Large-scale molecular dynamics simulations reveal that the spatial arrangement of Zr ions is a decisive parameter governing local polarization enhancement without requiring macroscopic lattice strain. Rather than a volume-driven mechanism, this enhancement stems directly from a local structural phase redistribution (rhombohedral, orthorhombic, and tetragonal). Chemical segregation isolates larger, continuous Ti-rich regions that expand highly polar rhombohedral domains and increase the dipolar correlation length, effectively shifting local phase stability toward higher temperatures.

In the relaxor regime ($x = 0.4$), this spatial clustering reinforces local polar order while preventing long-range ferroelectric alignment, resulting in enhanced dielectric permittivity, broader polarization decay, and a highly diffuse phase transition. Independent evaluations of characteristic thermal milestones—including the maximum permittivity and Burns temperatures—confirm that Zr clustering thermally stabilizes polar correlations. Microscopically, this reinforced local order accounts for the observed electrical hardening, characterized by elevated coercive fields and remanent polarization. This behavior is rooted in the formation of resilient, vortex-like topological states—both cluster-centered and interstitial—that act as pinning centers against polarization reversal. Importantly, while segregation maximizes the density and stability of these swirling textures, their fundamental origin lies in the intrinsic Zr–Ti off-centering mismatch, which remains active at the atomic scale even in homogeneous systems.

While these atomistic simulations evaluate idealized structural extremes within the intrinsic timescale limits of molecular dynamics, they establish a clear link between nanoscale compositional heterogeneity and macroscopic functional performance. Ultimately, these insights demonstrate that tailoring local chemical order offers a powerful strategy for controlling nanoscale topological states and optimizing lead-free perovskites for advanced functional applications.

**Declaration of generative AI and AI-assisted technologies in the writing process**

During the preparation of this work, the authors used Google Gemini in order to improve spelling and readability. After using this tool, the authors carefully reviewed and edited the content word by word and take full responsibility for the content of the publication.

**CRediT authorship contribution statement**

**Matias Baldassin:** Methodology – Visualization – Software – Validation – Formal analysis – Investigation – Data Curation Writing - Review & Editing. **Rodrigo Machado:** Conceptualization – Software – Validation – Investigation – Data Curation – Writing - Review & Editing. **Marcelo Sepliarsky:** Methodology – Software Validation – Investigation – Writing - Review & Editing. **Marcelo Stachiotti:** Conceptualization – Validation – Investigation – Writing – Original Draft – Supervision.

## Declaration of competing interest

The authors declare that they have no known competing financial interests or personal relationships that could have appeared to influence the work reported in this paper.

## Data availability

All data are in the main text or the Supplementary Material. More details and access to the raw and processed data will be made available on request.

## Acknowledgments

This work was supported by Consejo Nacional de Investigaciones Científicas y Técnicas de la República Argentina (CONICET) by PIP-0374. MGS thanks support from Consejo de Investigaciones de la Universidad Nacional de Rosario (CIUNR). We acknowledge the financial support of the Project 101236483-3D-TOPO-HORIZON-MSCA-2024-SE-01.

## Supplementary materials

Supplementary material associated with this article can be found, in the online version, at ...